\documentclass[prl,twocolumn,amsmath,amssymb,floatfix,showpacs,superscriptaddress]{revtex4}

\usepackage{graphicx}

\usepackage{amssymb}
\usepackage{color}
\usepackage{epsfig}

\begin{document}
\def\rhov{{\mbox{\boldmath{$\rho$}}}}
\def\tauv{{\mbox{\boldmath{$\tau$}}}}
\def\Lambdav{{\mbox{\boldmath{$\Lambda$}}}}
\def\sigmav{{\mbox{\boldmath{$\sigma$}}}}
\def\xiv{{\mbox{\boldmath{$\xi$}}}}
\def\chiv{{\mbox{\boldmath{$\chi$}}}}
\def\oh{{\scriptsize 1 \over \scriptsize 2}}
\def\ot{{\scriptsize 1 \over \scriptsize 3}}
\def\of{{\scriptsize 1 \over \scriptsize 4}}
\def\tf{{\scriptsize 3 \over \scriptsize 4}}
\title{Order Parameters and Phase Diagram of Multiferroic RMn$_2$O$_5$}

\author{A. B. Harris}

\affiliation{Department of Physics and Astronomy, University of
Pennsylvania, Philadelphia, PA 19104}

\author{Amnon Aharony}

\altaffiliation{Also emeritus, Tel Aviv University.}

\affiliation{Department of Physics,
Ben Gurion University, Beer
Sheva 84105 ISRAEL}

\author{Ora Entin-Wohlman}

\altaffiliation{Also emeritus, Tel Aviv University.}

\affiliation{Department of Physics,
Ben Gurion University, Beer Sheva 84105 ISRAEL}
\date{\today}

\begin{abstract}
The generic magnetic phase diagram of multiferroic RMn$_2$O$_5$
(with R=Y, Ho, Tb, Er, Tm), which allows different sequences of
ordered magnetic structures for different R's and different
control parameters, is described using order parameters which
explicitly incorporate the magnetic symmetry. A phenomenological
magneto-electric coupling is used to explain why some of these
magnetic phases are also ferroelectric. Several new experiments,
which can test this theory, are proposed.
\end{abstract}
\maketitle

There has been much recent interest in multiferroics, which
display simultaneous magnetic and ferroelectric (FE) ordering
\cite {KIMURA,CHAPON1,LAWES,MK}. In particular, the orthorhombic
family RMn$_2$O$_5$ (RMO), where R is a rare earth, exhibits
interesting sequences of magnetic density wave orderings, with
varying wave vector ${\bf q}$, and some of these phases are also
FE \cite{CHAPON1,NODA,koba_Er,koba_Tb,koba_Tm,KIMURA6}. In all
these phases one has $q_y=0$, while $|q_x-{1\over 2}|\lesssim
0.02$ and $|q_z-{1\over 4}|\lesssim 0.02$. Cooling from the
paramagnetic (PM) phase, one first enters a phase in which both
$q_x$ and $q_z$ are incommensurate. We call this phase II$_1$
(I=``incommensurate", and the subscript will be explained below;
some experimental papers call this phase 2DIC). For R=Y
\cite{NODA}, Er \cite{koba_Er} and Tm \cite{koba_Tm}, further
cooling yields transitions into a phase which we call IC$_2$ (also
called 1DIC), where $q_x$ is still incommensurate, while $q_z={1
\over 4}$ (C=``commensurate"), then into a ``CC" phase (also
called CM), with ${\bf q}=({1 \over 2},~0,~{1 \over 4})$, and
finally into a phase where both $q_x$ and $q_z$ are incommensurate
again (``II$_2$", or LTI-2DIC). R=Ho \cite{KIMURA6} and Tb
\cite{koba_Tb} go directly from II$_1$ to CC. For R=Er, the low
temperature ($T$) phase has $q_x={1 \over 2}$, while  $q_z$ is
incommensurate (``CI", or LTI-1DIC).  While the phases IC$_2$ and
CC exhibit a FE moment ${\bf P}$ along the $y~(b)$ axis, such a
moment appears in only some of the observed low $T$ phases
\cite{KAGOM,HIGA05}.
 Up to now, the microscopic theories of these systems are
 controversial, and a phenomenological description which provides a unified
explanation of this complicated phase behavior does not exist. The
present paper rectifies this situation, and provides a basis for
analyses of other multiferroics with large unit cells.

\begin{figure}[ht]
\begin{center}
\hspace{-0.1 in}
\hspace{4cm}
\hspace{0.05 in}
\includegraphics[width=4.3 cm]{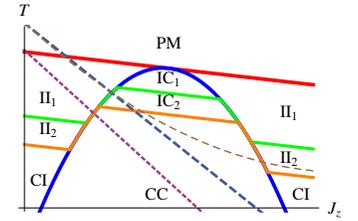}
\vspace{-0.7cm}
\end{center}
\caption{LHS (supplied separately): Schematic 3D phase
diagram for $q_x$ ($q_z$) near ${1 \over 2}$ (${1 \over 4}$). $J_x$ and
$J_z$ are  parameters which control $q_x$ and $q_z$. 
The red surface separates PM and II$_1$. Below the blue surface
one has $q_z={1 \over 4}$, in phases IC$_1$ and IC$_2$. The green
surfaces represent II$_1$$\rightarrow$II$_2$ and
IC$_1$$\rightarrow$IC$_2$. Below the orange surfaces $q_x={1 \over
2}$, in phases CI or CC. RHS: a cut at constant $q_x \not=1/2$.
The dashed and dotted lines represent proposed trajectories for
specific RMO's, as $T$ is varied.}\label{fig1}
\end{figure}

Although group theory has been applied to neutron diffraction data
from magnetic materials \cite{ref}, its implications for
multiferroics have not been fully exploited until the definitive
analyses of  Ni$_3$V$_2$O$_8$ and TbMnO$_3$
\cite{LAWES,MK,PRB,ABH,RC}. Following the same approach, we
identify the order parameters (OP's) allowed by symmetry
\cite{others} 
and find the generic phase diagram for RMO systems (Fig.
\ref{fig1}), which allows for the observed sequences of phases.
The theory also explains (a) which phases are simultaneously
magnetic and ferroelectric, (b) the occurrence of two distinct
spin structures in neutron diffraction studies of the CC phase
\cite{CHAPON,NODA}, and makes several new predictions, which can
be tested experimentally.

The PM  unit cell of the RMO's contains 4 Mn$^{3+}$, 4 Mn$^{4+}$
and 4 R$^{3+}$ ions. Denoting these ions  by $\tau(=1, \dots,12$),
and the corresponding Fourier transforms of the
$\alpha$-spin-components by $S_\alpha({\bf q},\tau)$, the
quadratic terms in the Landau free energy $F_M$ are
\begin{eqnarray}
F_{M,2}={1 \over 2}\sum_{{\bf
q},\alpha,\beta;\tau,\tau'}\chi_{\alpha\beta}^{-1} ({\bf
q};\tau,\tau')S_\alpha({\bf q},\tau)^\ast S_\beta({\bf q},\tau').
\end{eqnarray}
In principle one would diagonalize the ($36 \times 36$) inverse
susceptibility matrix $\chi_{\alpha\beta}^{-1}({\bf
q};\tau,\tau')$ (determined by the various magnetic interactions).
As $T$ is lowered, the first phase to order corresponds to the
eigenvalue which approaches zero first. The degeneracy of this
eigenvalue has two origins: first, all the $n_q$ wave vectors in
the star of symmetry-related optimalwave vectors ${\bf q}$'s have the same eigenvalue.
Second, each of these ${\bf q}$'s is associated with irreducible
representations (irrep's) $\Gamma$ of the PM symmetry group of
{\bf q} (the `little group') \cite{LL}. Excepting accidental
degeneracy, a continuous transition from the PM phase involves
only a single irrep. If this critical irrep is $d$ dimensional
(dD), then this eigenvalue is $dn_q$-fold degenerate and this
manifold is described by $dn_q$ real OP's, or $dn_q/2$ complex
ones. Each complex OP represents the amplitude and the phase of
the spin ordering eigenfunction, $\{S_\alpha({\bf q},\tau)\}$. The
symmetry of the eigenfunction is associated with the irrep and is
inherited by the OP's.

For each RMO, the optimal wave vector {\bf q} is determined by its
specific material (e. g. the exchange and anisotropy energies) and
experimental (e. g. pressure, magnetic field) parameters. We
represent these control parameters by their combinations, denoted
$J_x$ and $J_z$, which fix the values of $q_x$ and $q_z$,
respectively.  Figure \ref{fig1} shows the phase diagram of the
RMO's in terms of $J_x$ and $J_z$. Following experiments, we fix
$q_y=0$. We start with the case $q_x\ne {1 \over 2}$ (with $q_z$
near ${1\over 4}$). For each such ${\bf q}$, the `little' group
contains only unity and $m_y$, which maps $(x,y,z)$ into $(x+{1
\over 2},{\bar y}+{1 \over 2},z)$. This group has two 1D irreps,
$\Gamma_a$ and $\Gamma_b$, with complex OP's $\sigma^{}_a({\bf
q})$ and $\sigma^{}_b({\bf q})$. Inversion symmetry ${\cal I}$
then implies non-trivial relations between the $S_\alpha({\bf
q})$'s and the $S_\beta(-{\bf q})$'s (which have the same
eigenvalue), reducing the number of independent parameters. This
should ease future accurate analyses of the neutron data.
For $\Gamma_a$, symmetry
implies
\begin{eqnarray}
m_y \sigma^{}_a &=& \lambda_a^* \sigma^{}_a \ , \
\ {\cal I} \sigma^{}_a = e^{i \rho}  \sigma_a^*\ ,
\label{eq3}
\end{eqnarray}
and similarly for $\Gamma_b$ ($\rho$ depends on the origin)
\cite{inv}.

For $q_x \ne {1 \over 2}$, the star of ${\bf q}$ contains four
wave vectors, namely, ${\bf q}_\pm = \bigl (q_x, 0, \pm q_z
\bigr )$ and $-{\bf q}_\pm$. 
Therefore, we introduce two complex OP's, $\sigma^+_a \equiv
\sigma^{}_a({\bf q}_+^{(a)})$ and $\sigma^-_a \equiv
\sigma^{}_a({\bf q}_-^{(a)})$, associated with irrep $\Gamma_a$
and similarly for $\Gamma_b$.  Then,
$F_{M,2} = \sum_{s=a,b} (T-T_{C,s}) [ |\sigma^{+}_s|^2
+ |\sigma^{-}_s|^2]$.
Rejecting accidental degeneracy, we set $T_{C,a} > T_{C,b}$ and
identify the 2DIC phase with our II$_1$ phase, associated with a single irrep (the subscript 1
refers to the number of irreps), represented by the
$\sigma^{\pm}_a$'s. The transition PM $\rightarrow$ II$_1$ occurs
at $T=T_{C,a}$, represented by the top (red) surface in Fig.
\ref{fig1}. Which OP's actually order depends on the quartic terms
in the free energy. For $q_x\ne {1 \over 2}$, these include
\begin{eqnarray}
&&F_{M,4}^{(a)}= V_a (|\sigma^{+}_a|^2 + |\sigma^{-}_a|^2)^2 +U_a
|\sigma^{+}_a \sigma^{-}_a|^2
\nonumber\\
&&+\Sigma_{\bf G}\bigl  (W_{aa} \bigl [\sigma^{+}_a
(\sigma^{-}_a)^\ast\bigr ]^2 + cc \bigr )\delta \bigl [{\bf
G}-(0,0,4q_z)\bigr ], \label{2DIC}
\end{eqnarray}
where ${\bf G}$ is a reciprocal lattice vector. For $q_z\ne {1
\over 4}$ and $T<T_{C,a}$ one has $|\sigma_a^+|=|\sigma_a^-|>0$ if
$U_a<0$, and only one of the OP's orders otherwise.  For $q_z$
near ${1 \over 4}$, the {\it Umklapp} term with $W_{aa}$ locks
$q_z$ to ${1 \over 4}$, in a phase called IC$_1$. Within Landau
theory, this happens below a first order surface
(blue in Fig. \ref{fig1}), parabolic in $J_z$.

As $T$ is reduced, more quartic terms need to be considered,
notably $W \sum_{m=\pm} \bigl \{ [ \sigma^{}_a ({\bf q}_m^{(a)})
\sigma^{}_b({\bf q}_m^{(b)})^*]^2 + {\rm cc} \bigr \} \delta\bigl
({\bf q}_m^{(a)}-{\bf q}_m^{(b)}\bigr ) $. Assuming
 that ${\bf q}_\pm^{(a)}$
and ${\bf q}_\pm^{(b)}$ are almost the same, this term locks the
optimal ${\bf q}_\pm^{(b)}$ to ${\bf q}_\pm^{(a)}$, at some $T$
slightly below $T_{C,a}$, where $\sigma^{}_b$ has not yet ordered (without involving a phase transition.)
Accordingly, we no longer keep the superscripts $(a,b)$ on the
${\bf q}$'s. As $T$ is further reduced, the tendency of the spins
to have fixed length (rather than oscillate sinusoidally)
\cite{NAG,LAWES} may cause a second continuous transition, into
the phase II$_2$ (or IC$_2$), where both $\sigma^{}_a$ and
$\sigma^{}_b$ are nonzero. As shown below, this transition (green
surface in Fig. \ref{fig1}) occurs at a temperature which is
parabolic in $J_x$.

We next discuss the special case $q_x={1\over 2}$ (or
$J_x=J_{x,c}$, at the back of the 3D diagram in Fig. \ref{fig1}).
For ${\bf q}=({1 \over 2},0,q_z)$, the little symmetry group
changes: it now contains the additional glide operation $m_x$
[which maps $(x,y,z)$ into $(-x+{1 \over 2},y+{1 \over 2},z)$].
This group has only one 2D irrep \cite{BLAKE,ABH}, with two
degenerate complex OP's, $\sigma^{}_1$ and $\sigma^{}_2$, and
corresponding eigenvectors as listed in Table XVI of Ref.
\onlinecite{ABH} \cite{err}. These OP's transform as \cite{ABH}
\begin{eqnarray}
m_x \sigma^{}_n = \zeta_n \sigma^{}_n,\ \ \  m_y
\sigma^{}_n=\zeta_n \sigma^{}_{3-n},\ \ \
{\cal I} \sigma^{}_n = \sigma^*_{3-n}, \label{SYMEQ}
\end{eqnarray}
where $\zeta_n \equiv (-1)^{3-n},~n=1,2$. Cooling from the PM
phase, exactly at $J_x =J_{x,c}$ and $q_z \ne {1 \over 4}$, one first goes
into the CI phase, with the free energy
\begin{eqnarray}
F_M &=& (T-T_C) [|\sigma^{}_1 |^2
+|\sigma^{}_2 |^2]+u [|\sigma^{}_1 |^2 +|\sigma^{}_2 |^2]^2 \nonumber\\
&+& W_C |\sigma^{}_1 \sigma^{}_2 |^2 + V_C [\sigma^{}_1 \sigma_2^*
+ \sigma^{}_2 \sigma_1^* ]^2 . \label{F6}
\end{eqnarray}
On further cooling, additional {\it Umklapp} terms
 cause a first order transition into the CC phase where ${\bf q}=({1 \over 2},0,{1 \over 4})$. This lock-in
happens under a parabola, which connects to the blue parabolas
which appear for $q_x\ne{1\over 2}$.

We next vary $J_x$ away from $J_{x,c}$. The RHS of Fig. \ref{fig1}
shows a cut of the 3D phase diagram, at fixed $J_x=J_{x,c}+\Delta
J$. For small $\Delta J$, mirror symmetry,  $q_x \rightarrow -
q_x$, implies that the inverse susceptibility has two branches of
eigenvalues given by
$\chi^{-1}_\pm (q_x) \cong T-T_C + a k_x^2 \pm b k_x \Delta J$,
where $k_x=1/2-q_x$ and $a$ and $b$ are constants. At quadratic
order ({\it i. e.} using $F_{M,2}$), this implies that
$T_{C,a}-T_{C,b} = 2b k_x^{(0)} \Delta J$, where $k_x^{(0)}= b
\Delta J /(2a)$ minimizes $\chi^{-1}_\pm (q_x)$. This gives rise
to the (green) II$_1\rightarrow$II$_2$ phase boundary,
$T_{C,a}-T_{C,b} \propto (\Delta J)^2$.  

For $q_x$ close to ${1\over 2}$, further cooling may lock it to
${1\over 2}$, due to {\it Umklapp} terms. This may generate
transitions into the CI or the CC phase, for $T$'s below the
orange surfaces in Fig. \ref{fig1}. A detailed analysis shows that
these surfaces are also parabolic in $\Delta J$. The actual
sequence of transitions then depends on which parabola is
narrower. In Fig. \ref{fig1} we show the case when the green
parabolas are broader than the orange ones. In this case, the
orange surface represents II$_2$$\rightarrow$CI and
IC$_2$$\rightarrow$CC. In the opposite case, the phases II$_2$ and
IC$_2$ never appear. As shown in Fig. \ref{fig1}, both parabolas
are shifted upwards below the blue surface, where $q_z={1\over
4}$, due to {\it Umklapp} terms.

\begin{figure}[ht]
\begin{center}
\includegraphics[width=7 cm]{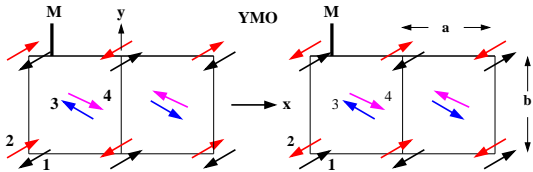}
\vspace{-.5cm}
\end{center}
\caption{(Color online) \label{YMOFIG} Schematic diagram of the ${\bf a}$
and ${\bf b}$ components of the Mn$^{3+}$ spins in a single a-b plane
of YMO for the CC phase. 
The glide $m_x$ consists of a mirror plane M at $x=a/4$ followed
by a translation b/2 along $y$.
LHS: the structure given in Table III of Ref.
\protect{\onlinecite{NODA}} (with the ${\bf c}$-components not
shown). RHS: the structure given in Fig. 2
[\protect{\onlinecite{TOP}}] of Ref. \protect{\onlinecite{CHAPON}}
(who reported zero \protect{${\bf c}$}-components of spin).}
\end{figure}

Equations (\ref{SYMEQ}-\ref{F6}) lead to a natural interpretation
of neutron scattering results for the CC phase in YMO. Figure
\ref{YMOFIG} shows the Mn$^{3+}$ {\bf a}-{\bf b} plane spin
components in the CC phase of YMO, from the data of Refs.
\onlinecite{NODA} \cite{NB1} and \onlinecite{CHAPON}. These two
structures are obviously similar, and one might ask what symmetry
(if any) relates them \cite{WANG}. Since the structure on the left
(right) is even (odd) under the glide operation $m_x$, we conclude
that the structure on the left (right) has $\sigma^{}_2=0$
($\sigma^{}_1=0$). Going between these two structures corresponds
to a rotation in OP space; the in-plane spin components belong to
distinct but equivalent structures. Since either $\sigma^{}_1=0$
or $\sigma^{}_2=0$, we conclude that in Eq. (\ref{F6}), the net
coeficient of $|\sigma^{}_1\sigma^{}_2|^2$ ($W_C - 4|V_C|$ plus
the additional {\it Umklapp} terms) is positive, preventing both
OP's from ordering simultaneously \cite{ABH}.

\vspace{-0.05 in}

For the phenomenological description of the multiferroicity, the
total free energy is $F_{\rm ME}= F_{\rm M} + {1 \over 2} {\bf
P}^2 \epsilon^{-1} + V_{\rm int}$, where $\epsilon$ is the
dielectric susceptibility and $V_{\rm int}$ the magnetoelectric
(ME) interaction. Although such interactions can also generate a
spatially non-uniform
 ${\bf P}$, here we
discuss only the uniform case. We again start with the II and IC
phases, where $q_x\ne {1 \over 2}$. To lowest-order, wave vector
conservation and time-reversal invariance give \cite{LAWES,goshen}
\begin{eqnarray}
V_{\rm int} &=& \sum_{s,t=a,b} \sum_{{\bf q}={\bf q}_\pm}
\sum_\gamma c^{}_{st\gamma} \sigma^{}_s({\bf q}) \sigma^{}_t(-{\bf
q}) P_\gamma. \label{EQ5} \end{eqnarray} The terms with $s=t$
vanish because they are odd under ${\cal I}$. For the II$_1$
(IC$_1$) phase, only $\sigma^{}_a$ is nonzero, and therefore ${\bf
P}=0$. To have ${\bf P} \ne 0$ with $q_x\ne {1\over 2}$ we must
have the superposition of two irreps, and this happens only in the
II$_2$ or the IC$_2$ phases. In these phases, we have
\begin{eqnarray}
V_{\rm int} &=& \sum_{{\bf q}={\bf q}_\pm}\sum_\gamma \bigl [
ir_\gamma \sigma^{}_a({\bf q}) \sigma_b(-{\bf q}) + cc \bigr ]
P_\gamma \ , \label{Vint}
\end{eqnarray}
and invariance under ${\cal I}$ requires that $r_\gamma$ is real.
From Eq. (\ref{eq3}), $\sigma^{}_a \sigma_b^*$ is odd under $m_y$.
For $V_{\rm int}$ to be invariant under $m_y$, $P_\gamma$ must be
odd under $m_y$: symmetry forces ${\bf P}$ to be along  ${\bf
y(b)}$, as observed (Higher order ME interactions weakly
violate this result \cite{inv}).
.

In the CC phase,  Eq. (\ref{EQ5}) is invariant under the symmetry
operations of Eq. (\ref{SYMEQ}) only if  [\onlinecite{ABH}]
\begin{eqnarray}
V_{\rm int} &=& {\rm const.} \times [ |\sigma^{}_1({\bf q})|^2 -
|\sigma^{}_2({\bf q})|^2] P_y \ . \label{EQ6} \end{eqnarray} Note
that Eqs. (\ref{Vint}) and (\ref{EQ6}) apply 
whether the {\it microscopic} ME interactions are due to exchange
striction \cite{Yildirim}
or to charge ordering \cite{Khomskii}. Thus, 
{\bf P} must lie along ${\bf y(b)}$ also in the CC phase
\cite{higher}. Within mean field theory, $P_b$ is proportional to
$|\langle \sigma^2 \rangle|$, as is the intensity of the magnetic
Bragg peaks. This is confirmed \cite{RFMO} in RbFe(MoO$_4$)$_2$
[which is also described by Eq. (\ref{EQ6})] and also apparently
for ErMO by Ref. \onlinecite{Boden} \cite{AA}. Since the CC phase
is ferroelectric, the fourth order terms in Eq. (\ref{F6}) (plus
the {\it Umklapp} terms) must select $\sigma^{}_1 \sigma^{}_2=0$,
which we deduced from Fig. \ref{YMOFIG}. (The alternative would
imply $|\sigma^{}_1|=|\sigma^{}_2|$, hence ${\bf P}=0$.) In fact,
the selection of which OP, $\sigma^{}_1$ or $\sigma^{}_2$, is
nonzero is a result of broken symmetry. An electric field along
${\bf y}({\bf b})$ would order $P_y$, and then Eq. (\ref{EQ6})
would select either $\sigma^{}_1$ or $\sigma^{}_2$, depending on
the sign of the field. Therefore we suggest that the sample should
be cooled into the FE phase in the presence of a small electric
field along $y$. Depending on the sign of the electric field one
should get either the left-hand or the right-hand panel of Fig. 2
\cite{YAMA}.

Equation (\ref{Vint}) has further implications. First, near the
P$\rightarrow$II$_1$ transition, a leading fluctuation expansion
yields $\Delta\epsilon \propto \langle P_b^2\rangle\propto
|\langle \sigma_a^2\rangle \langle \sigma_b^2\rangle|$. Since only
$\sigma^{}_a$ becomes critical there, we expect singularities in
$\epsilon$ which behave as the energy ($|T-T_{C1}|^{1-\alpha}$)
and (for $T<T_{C1}$) as the square of the OP
($(T_{C1}-T)^{2\beta}$), but with $n=4$ exponents \cite{RG}.
Indeed, experiments \cite{dela1} show a break in slope at
$T_{C1}$, apparently confirming this prediction. In addition, this
anomaly in the zero frequency dielectric function
$\epsilon(\omega=0)$ reflects the emergence of a resonance in
$\epsilon(\omega)$, due to electromagnons \cite{PIMENOV}.  Second,
in the II$_1$ phase $\langle\sigma^{}_a\rangle \ne 0$, so that
$V_{\rm int}$ becomes $-2r_b\Im[\langle \sigma^{}_a \rangle
\sigma_b^*] P_y$. This bilinear coupling between $P_y$ and
$\Im[\langle \sigma^{}_a \rangle \sigma_b^*]$ has several
implications on the critical behavior near the
II$_1$$\rightarrow$II$_2$, should this transition  be discovered
in some new RMO \cite{JPC}.

Finally, we associate the different RMO's with trajectories on our
phase diagram. Since  ErMO \cite{koba_Er}, TmMO \cite{koba_Tm},
and YMO \cite{NODA} exhibit ferroelectricity in the phase denoted
1DIC, we must identify this phase with our IC$_2$ phase, where
both $\sigma^{}_a$ and $\sigma^{}_b$ order. For these materials,
the experimental path in parameter space apparently goes from
II$_1$ via IC$_2$ into the CC phase, as indicated by the dashed
lines in the RHS of Fig. \ref{fig1} (these lines have small
slopes, since the experimental optimal ${\bf q}$ varies with $T$:
$J_x$ and $J_z$ depend on $T$ due to other degrees of freedom).
The last term in Eq. (\ref{2DIC}) implies that $q_z$ locks to ${1
\over 4}$ only if both ${\bf q}_+$ and ${\bf q}_-$ appear in the
IC$_1$ and IC$_2$ phases. If $2|W_{aa}|> U_a > 0$, then the two
${\bf q}$'s first appear as the IC$_1$ phase is entered. If $U_a
<0$, then both wave vectors would have already condensed
simultaneously in the II$_1$ phase.
It would be interesting to 
determine which scenario actually occurs. Since the ME interaction
is significant, we suggest to apply an electric field parallel to
one of the ${\bf q}$'s, and check whether in the II$_1$ phase the
two ${\bf q}$'s arise in separate domains or coexist within a
single domain, following the logic of Ref. \onlinecite{LYNN}. In
contrast to the above RMO's, HoMO \cite{KIMURA6} or TbMO
\cite{koba_Tb} go directly from 2DIC to CM, as along the dotted
line in the RHS of Fig. \ref{fig1}. 
Both sequences are thus allowed by our theory.

The Landau theory is probably less useful at lower $T$: the low
$T$ phases depend on the details of the magnetic interactions, and
higher order terms in $F_M$ should be included. Such terms could
turn the (orange or blue) surface bounding the CC phase backwards,
thus allowing transitions back into the paraelectric II$_1$ phase,
the weakly FE phases II$_2$ or IC$_2$ or the FE phase CI. Also,
the trajectory describing each material need not be straight
(thick dashed line in Fig. \ref{fig1}). A parabolic line, like the
thin dashed line, would yield a transition from CC to II$_2$ (or
even to II$_1$) with decreasing $T$. In fact, in ErMO
\cite{koba_Er} the LTI phase seems to have $q_x={1 \over 2}$,
which identifies this phase with our CI phase. Thus, the observed
LTI phase could be any of the phases on the other side of the CC
region, paraelectric or weakly ferroelectric. The effects of a
magnetic field can be explained as follows: the field generates
magnetic moments on the R ion (even above their ordering
temperature). Since these ions couple to the Mn ions, their moment
results in changes in the effective Mn-Mn interactions, thus
changing the `control parameters' and the optimal {\bf q}.
Apparently, this often moves the material towards the CM regime,
resulting in a transition from the low $T$ phase (II$_1$ or
II$_2$) back into the CC phase \cite{HIGA05,KIMURA6}. Similar
effects happen due to pressure \cite{dela1}. Neutron diffraction
measurements in a magnetic field and pressure could help resolve
these scenarios.

In summary: we have developed a phase diagram to explain the
multiferroic behavior of the family of RMO systems and have
proposed several experiments to explore the unusual symmetries of
these systems.

We thank  M. Kenzelmann, S. H. Lee and D. Mukamel for helpful interactions. AA
and OEW acknowledge support from the ISF and from the GIF, and the hospitality of
KITP, where this research was supported in part by the National
Science Foundation under Grant No. PHY05-51164.


\begin{thebibliography} {99}
\bibitem{KIMURA} 
T. Kimura {\it et al.}, Nature {\bf 426}, 55 (2003).
\bibitem{CHAPON1} 
L. C. Chapon {\it et al.}, Phys. Rev. Lett. {\bf 93}, 177402 (2004).
\bibitem{LAWES} 
G. Lawes {\it et al.}, Phys. Rev. Lett. {\bf 95}, 087205 (2005).
\bibitem{MK} 
M. Kenzelmann {\it et al.}, Phys. Rev. Lett. {\bf 95}, 087206
(2005).
\bibitem{NODA}
H. Kimura {\it et al.}, J. Phys. Soc. Jpn. {\bf 76}, 074706
(2007).
\bibitem{koba_Er} 
S. Kobayashi {\it et al.}, J. Phys. Soc. Jpn. {\bf 73}, 1031 (2004).
\bibitem{koba_Tb} 
S. Kobayashi {\it et al.}, J. Phys. Soc. Jpn. {\bf 73}, 3439 (2004).
\bibitem{koba_Tm} 
S. Kobayashi {\it et al.}, J. Phys. Soc. Jpn. {\bf 74}, 468 (2005).
\bibitem{KIMURA6} 
 H. Kimura {\it et al.}, J. Phys. Soc. Jpn. {\bf 75}, 113701 (2006).
\bibitem{KAGOM} 
I. Kagomiya {\it et al.}, Ferroelectrics {\bf 286}, 167 (2003).

\bibitem{HIGA05} 
D. Higashiyama {\it et al.}, Phys. Rev. B {\bf 72}, 064421 (2005).

\bibitem{ref} A. P. Cracknell, J. Phys. C. {\bf 4}, 2488 (1971); D. B. Litvin
and W. Opechowski, Physica {\bf 76}, 538 (1974); Yu. A. Izyumov,
V. E. Naish and R. P. Ozerov, {\it Neutron Diffraction of Magnetic
Materials} (Springer-Verlag, Amsterdam, 1991).


\bibitem{PRB} 
M. Kenzelmann {\it et al.}, Phys. Rev. B {\bf 74}, 014429 (2006).
\bibitem{ABH} 
A didactic review of the approach, with more examples, is given by A. B. Harris, Phys. Rev. B {\bf 76}, 054447 (2007).
\bibitem{RC} 
See also P. G. Radaelli and L. C. Chapon, Phys. Rev. B {\bf 76},
054428 (2007).
\bibitem{others} A simple theoretical description of multiferroics [M. Mostovoy, Phys. Rev. Lett. {\bf 96}, 067601 (2006);
J. J. Betouras {\it et al.}, Phys. Rev. Lett.  {\bf 98}, 257602
(2007)] uses a {\it single} wave vector {\bf q} and a {\it single}
vector spin amplitude, ${\bf S}({\bf q})$, and constructs
combinations of ${\bf S}({\bf q})$, ${\bf S}(-{\bf q})$, {\bf q}
and ${\bf P}$ allowed by symmetry. However, it is not clear how to
relate the single ${\bf S}({\bf q})$ to the different sublattices
within the large unit cell of RMO [see also Ref. \onlinecite{RFMO}
and M. Kenzelmann and A. B. Harris, Phys. Rev. Lett. {\bf 100},
089701 (2008).]


\bibitem{RFMO} 
M. Kenzelmann {\it et al.}, Phys. Rev. Lett. {\bf 98}, 267205
(2007).


\bibitem{CHAPON} 
L. C. Chapon {\it et al.}, Phys. Rev. Lett. {\bf 96}, 097601 (2006).
\bibitem{LL} 
L. D. Landau and I. M. Lifshitz, {\it Statistical Physics}
(Pergamon 1978), Sec. 139.

\bibitem{inv} A. B. Harris, M. Kenzelmann, A. Aharony, and O.
Entin-Wohlman, Phys. Rev. B (in press); arXiv:0803.0945.




\bibitem{NAG} 
T. A. Kaplan, Phys. Rev. {\bf 124}, 329 (1961); T. Nagamiya, in
{\it Solid State Physics}, ed. F. Seitz and D. Turnbull (Academic,
New York, 1967), Vol. 20, p. 346.

\bibitem{BLAKE} 
G. R. Blake {\it et al.}, Phys. Rev. B {\bf 71}, 214402 (2005).
\bibitem{err} See also A. B. Harris, Phys. Rev. B {\bf 77}, 019901(E)
(2008).

\bibitem{TOP} 
The labeling of the top and bottom panels in Ref.
\onlinecite{CHAPON} seems to be switched.
\bibitem{NB1} 
The wave function found in Ref. \onlinecite{NODA} (which allowed
free variation of all the spins in the unit cell)
agrees closely with the eigenvector corresponding to the full
group theory predictions, as listed in Table XVI of Ref.
\onlinecite{ABH}.
\bibitem{WANG} 
This degeneracy was also found in a first-principles calculation:
C. Wang {\it et al.}, Phys. Rev. Lett. {\bf 99}, 177202 (2007).
\bibitem{goshen} This phenomenological description is similar to
that used for simpler commensurate antiferromagnets by S. Goshen
{\it et al.}, Phys. Rev. B {\bf 2}, 4679 (1970).

\bibitem{Yildirim} A. B. Harris, T. Yildirim, A. Aharony, and O. Entin-Wohlman,
Phys. Rev. B {\bf 73}, 184433 (2006).

\bibitem{Khomskii}
J. van den Brink and D. Khomskii, arXiv:0803.2964.


\bibitem{higher} In the CC phase, higher order terms, similar to those of
I. A. Sergienko {\it et al.} [Phys. Rev. Lett. {\bf 97}, 227204 (2006)],
do not come into play here because the quartic terms in Eq. (\ref{F6})
favor either $\sigma_1 \sigma_2=0$ or $|\sigma_1|=|\sigma_2|$.
\bibitem{Boden} 
Y. Bodenthin {\it et al.}, cond-mat/0707.0180.
\bibitem{AA} 
However, critical fluctuations imply different exponents for $P_b$
and $|\sigma|^2$ [A. Aharony {\it et al.}, Phys. Rev. Lett. {\bf
57}, 1012 (1986)], and this should be checked more carefully.
\bibitem{YAMA} 
A related experiment was recently
performed in TbMnO$_3$ [Y. Yamasaki {\it et al.}, Phys. Rev. Lett. {\bf 98}, 147204
(2007)]. Radaelli {\it et al.} also inform us that their unpublished experiments on YMO are consistent with our predictions.

\bibitem{RG} The isotropic $n=4$ fixed point is slightly unstable against
$U_a$. For $U_a>0$ there is probably a crossover to a weak first
order transition.


\bibitem{dela1} 
C. R. dela Cruz {\it et al.}, cond-mat/0707.0318.

\bibitem{PIMENOV} 
A. Pimenov {\it et al.}, Nature Phys. {\bf 2}, 97 (2006); A. B. Sushkov {\it et al.}, Phys. Rev. Lett. {\bf 98}, 027202 (2007).

\bibitem{JPC} A. B. Harris, A. Aharony and O. Entin-Wohlman,
J. Phys. C (in press); arXiv:0804.3039.

\bibitem{LYNN} 
S. Skanthakumar {\it et al.}, Phys. Rev. B {\bf 47}, 6173 (1993).

\end{thebibliography}
\end{document}